\documentclass[a4paper,english,aps, prb]{revtex4-1}
\usepackage[T1]{fontenc}
\usepackage[latin9]{inputenc}
\usepackage{babel}
\usepackage{amsmath}
\usepackage{amssymb}
\usepackage{graphicx}
\usepackage[unicode=true,pdfusetitle,
 bookmarks=true,bookmarksnumbered=false,bookmarksopen=false,
 breaklinks=false,pdfborder={0 0 1},backref=false,colorlinks=false]
 {hyperref}

\makeatletter

\providecommand{\tabularnewline}{\\}

\makeatother

\begin{document}
\title{Gate-tunable cross-plane heat dissipation in single-layer transition
metal dichalcogenides}
\author{Zhun-Yong Ong}
\email{ongzy@ihpc.a-star.edu.sg}

\affiliation{Institute of High Performance Computing, A{*}STAR, Singapore 138632,
Singapore}
\author{Gang Zhang}
\affiliation{Institute of High Performance Computing, A{*}STAR, Singapore 138632,
Singapore}
\author{Linyou Cao}
\affiliation{Department of Materials Science and Engineering, North Carolina State
University, Raleigh, North Carolina, 27695, USA}
\affiliation{Department of Physics, North Carolina State University, Raleigh, North
Carolina, 27695, United States}
\affiliation{Department of Electrical and Computer Engineering, North Carolina
State University, Raleigh, North Carolina, 27695, USA}
\author{Yong-Wei Zhang}
\affiliation{Institute of High Performance Computing, A{*}STAR, Singapore 138632,
Singapore}
\date{\today}
\begin{abstract}
Efficient heat dissipation to the substrate is crucial for optimal
device performance in nanoelectronics. We develop a theory of electronic
thermal boundary conductance (TBC) mediated by remote phonon scattering
for the single-layer transition metal dichalcogenide (TMD) semiconductors
MoS$_{2}$ and WS$_{2}$, and model their electronic TBC with different
dielectric substrates (SiO$_{2}$, HfO$_{2}$ and Al$_{2}$O$_{3}$).
Our results indicate that the electronic TBC is strongly dependent
on the electron density, suggesting that it can be modulated by the
gate electrode in field-effect transistors, and this effect is most
pronounced with Al$_{2}$O$_{3}$. Our work paves the way for the
design of novel thermal devices with gate-tunable cross-plane heat-dissipative
properties.
\end{abstract}
\maketitle

\section{Introduction}

Atomically thin two-dimensional (2D) transition metal dichalcogenide
(TMD) semiconductors such as MoS$_{2}$ and WS$_{2}$ are promising
alternative materials for the development of next-generation electronic
devices~\citep{XLi:AdvMater15_Performance,DLembke:ACR15_Single}.
At the nanoscale, the high power densities in these devices require
efficient thermal management which is crucial for optimal device performance,
with the thermal boundary conductance (TBC) of the 2D crystal-substrate
interface playing a key role in the dissipation of excess Joule heat~\citep{EPop:NR10_Energy,ZYOng:2DM19_Energy}.
Therefore, clearer insights into the physical mechanisms underlying
heat dissipation across the TMD-substrate interface may lead to the
development of superior thermally aware TMD-based nanoelectronic designs
as well as novel applications in thermal energy harvesting or for
channeling heat in ultracompact geometries.

One widely studied mechanism is the van der Waals coupling between
the phonons of the 2D crystal and its substrate which is believed
to be the dominant component in the overall TBC~\citep{ZYOng:2DM19_Energy}.
The phononic TBC has been estimated using molecular dynamics (MD)
simulations~\citep{ZYOng:JAP18_Flexural,SVSuryavanshi:JAP19_Thermal},
elasticity theory~\citep{BNJPersson:JPCM11_Phononic,ZYOng:PRB16_Theory,ZYOng:PRB17_Thickness}
and density-functional-theory-based models~\citep{GCCorrea:Nanotech17_Interface}.
Another mechanism of heat dissipation is the inelastic scattering
of electrons in the 2D crystal by dipoles in the dielectric substrate,
a phenomenon known widely as ``remote phonon (RP) scattering'' or
``surface optical phonon scattering''~\citep{KHess:SSC79_Remote,MVFischetti:JAP01_Effective,AKonar:PRB10_Effect,KZou:PRL10_Deposition}.
This phenomenon occurs for an insulating metal-oxide dielectric substrate
(e.g. SiO$_{2}$) in which the oscillating dipoles originating from
the bulk polar optical phonons create an evanescent surface electric
field that scatters the electrons remotely. The associated electron-phonon
coupling strength depends on the dielectric properties of the substrate
and screening by the electrons in the TMD. In a high-$\kappa$ dielectric
substrate such as HfO$_{2}$ or Al$_{2}$O$_{3}$, inelastic RP scattering
can be strong enough to cause substantial momentum relaxation of the
TMD electrons~\citep{ZYu:AdvMater16_Realization}. In graphene~\citep{KZou:PRL10_Deposition,ZYOng:PRB13_Signatures},
this mechanism is also expected to facilitate energy dissipation from
the nonequilibrium electrons to the substrate.

In spite of its role in limiting the electron mobility in TMDs~\citep{ZYu:NatCommun14_Towards,ZYu:AdvMater16_Realization},
our understanding of heat dissipation through RP scattering remains
limited for TMDs. Unlike the phononic TBC which depends on interatomic
forces, the RP-mediated electronic TBC ($G_{\text{el}}$) is expected
to vary with the electron density ($n$) which can be modulated by
the gate electrode in a field-effect transistor. Thus, a deeper understanding
of this phenomenon may inspire the design of novel devices with gate-tunable
thermal properties and perhaps shed light on the distribution of reported
experimental TBC values~\citep{ATaube:ACSApplMaterInterf15_Temperature,EYalon:NL17_Energy,EYalon:ACSApplMaterInterf17_Temperature,PYasaei:AdvMaterInterf17_Interfacial}.
Although earlier theoretical and experimental work on RP heat dissipation
in graphene~\citep{ZYOng:PRB13_Signatures,YKoh:NL16_Role} indicates
that $G_{\text{el}}$ is small for the graphene-SiO$_{2}$ interface
and weakly dependent on the electron density, the different electronic
structure in single-layer MoS$_{2}$ and WS$_{2}$ suggests that these
findings for graphene may not apply to TMDs.

In our paper, we develop a theoretical model of heat dissipation by
RP scattering~\citep{ZYOng:PRB13_Signatures} and apply it to investigate
the electronic TBC and its dependence on the electron density, substrate
material (SiO$_{2}$, HfO$_{2}$, and Al$_{2}$O$_{3}$) and temperature
($T$) in two commonly studied single-layer TMDs (MoS$_{2}$ and WS$_{2}$).
The effects of electron screening on $G_{\text{el}}$ are studied.
One of our main findings is that the $G_{\text{el}}$ for MoS$_{2}$
and WS$_{2}$ exhibits a substantially greater dependence on $n$
than the $G_{\text{el}}$ for graphene, with $G_{\text{el}}$ reaching
$85$ MW/K/m$^{2}$ for the MoS$_{2}$-Al$_{2}$O$_{3}$ interface
at the electron density of $10^{13}$ cm$^{-2}$. We suggest applications
for different substrate materials and how this electron density dependence
of the TBC can be exploited to create gate-tunable thermal insulators.

\section{Theoretical model for heat dissipation by remote phonon scattering}

The key elements of our RP model are based on Ref.~\citep{ZYOng:PRB13_Signatures}.
We treat the electrons in the TMD as a single-band 2D electron gas
(2DEG) at a fixed distance $d$ above the substrate, which we approximate
as a dielectric continuum. The treatment of the TMD as a 2D electron
gas can be justified by the three-atom thickness of single-layer TMD
crystals, which implies that the electrons are strongly confined in
the out-of-plane direction. The 2DEG in the TMD has a parabolic dispersion
characterized by the effective mass $m_{e}$ with spin and valley
degeneracies $g_{s}$ and $g_{v}$, respectively. For the substrate,
the bulk polar optical phonons, of which there are typically two branches
for a dielectric insulator such as SiO$_{2}$, are characterized by
their longitudinal optical (LO) and transverse optical (TO) frequencies~\citep{MVFischetti:JAP01_Effective},
which are related through the equation $\epsilon_{\text{sub}}^{0}=\epsilon_{\text{sub}}^{\infty}\left(\frac{\omega_{\text{LO1}}^{2}}{\omega_{\text{TO1}}^{2}}\right)\left(\frac{\omega_{\text{LO2}}^{2}}{\omega_{\text{TO2}}^{2}}\right)$,
where $\epsilon_{\text{sub}}^{0}$ and $\epsilon_{\text{sub}}^{\infty}$
represent, respectively, the static and optical permittivity of the
substrate, and $\omega_{\text{LO1}}$ and $\omega_{\text{LO2}}$ ($\omega_{\text{TO1}}$
and $\omega_{\text{TO2}}$) are, respectively, the LO (TO) frequencies
of the first and second phonon branches with $\omega_{\text{LO1}}<\omega_{\text{LO2}}$
($\omega_{\text{TO1}}<\omega_{\text{TO2}}$). The LO phonon frequencies
are determined from the roots of the frequency-dependent substrate
dielectric function 
\begin{equation}
\epsilon_{\text{sub}}(\omega)=\epsilon_{\text{sub}}^{\infty}+(\epsilon_{\text{sub}}^{i}-\epsilon_{\text{sub}}^{\infty})\frac{\omega_{\text{TO2}}^{2}}{\omega^{2}-\omega_{\text{TO2}}^{2}}+(\epsilon_{\text{sub}}^{0}-\epsilon_{\text{sub}}^{i})\frac{\omega_{\text{TO1}}^{2}}{\omega^{2}-\omega_{\text{TO1}}^{2}}\label{eq:FreqDependDielectricFn}
\end{equation}
where $\omega$ and $\epsilon_{\text{sub}}^{i}$ are respectively
the frequency and intermediate permittivity of the substrate. We can
rewrite Eq.~(\ref{eq:FreqDependDielectricFn}) as $\epsilon_{\text{sub}}(\omega)=\epsilon_{\text{sub}}^{\infty}\left(\frac{\omega_{\text{LO1}}^{2}-\omega^{2}}{\omega_{\text{TO1}}^{2}-\omega^{2}}\right)\left(\frac{\omega_{\text{LO2}}^{2}-\omega^{2}}{\omega_{\text{TO2}}^{2}-\omega^{2}}\right)$.
The corresponding surface optical (SO) frequencies associated with
the surface electric field, $\omega_{\text{SO1}}$ and $\omega_{\text{SO2}}$
are determined from the solutions to $\epsilon_{\text{sub}}(\omega)+\epsilon_{0}=0$~\citep{AKonar:PRB10_Effect}.
The electronic TBC is determined from the rate of change of the SO1
and SO2 phonons as they scatter with the electrons in the TMD and
the resultant rate of energy dissipation from the electrons, following
the approach in Ref.~\citep{ZYOng:PRB13_Signatures}

\subsection{Surface optical phonon emission and absorption rates}

From first-order perturbation theory, the rate of change in the Bose-Einstein
(BE) occupation factor $N_{\gamma,\boldsymbol{q}}$ of the $\gamma$
phonon with wave vector $\boldsymbol{q}$, where $\gamma=$SO1 and
SO2 indexes the SO phonon branch, is~\citep{ZYOng:PRB13_Signatures}
\begin{equation}
\frac{dN_{\gamma,\boldsymbol{q}}}{dt}=-g_{s}g_{v}\sum_{\boldsymbol{p}}(W_{\gamma,\boldsymbol{p}\rightarrow\boldsymbol{p+q}}^{(\text{abs})}-W_{\gamma,\boldsymbol{p}\rightarrow\boldsymbol{p-q}}^{(\text{ems})})\ .\label{eq:PhononRate}
\end{equation}
The phonon absorption ($W_{\gamma,\boldsymbol{p}\rightarrow\boldsymbol{p+q}}^{(\text{abs})}$)
and emission ($W_{\gamma,\boldsymbol{p}\rightarrow\boldsymbol{p-q}}^{(\text{ems})}$)
terms describe phonon absorption and emission due to the $\boldsymbol{p}\rightarrow\boldsymbol{p+q}$
and $\boldsymbol{p}\rightarrow\boldsymbol{p-q}$ electronic transitions,
respectively, where $\boldsymbol{p}$ is the wave vector of the initial
electronic state. The expressions for $W_{\gamma,\boldsymbol{p}\rightarrow\boldsymbol{p+q}}^{(\text{abs})}$
and $W_{\gamma,\boldsymbol{p}\rightarrow\boldsymbol{p-q}}^{(\text{ems})}$
can be derived from the Fermi golden rule to yield 
\begin{equation}
\left\{ \begin{array}{c}
W_{\gamma,\boldsymbol{p}\rightarrow\boldsymbol{p+q}}^{(\text{abs})}\\
W_{\gamma,\boldsymbol{p}\rightarrow\boldsymbol{p-q}}^{(\text{ems})}
\end{array}\right\} =\frac{2\pi|M_{\gamma,\boldsymbol{q}}|^{2}}{\hbar}\left\{ \begin{array}{c}
f_{\boldsymbol{p}}(1-f_{\boldsymbol{p+q}})N_{\gamma,\boldsymbol{q}}\delta(E_{\boldsymbol{p+q}}-E_{\boldsymbol{p}}-\hbar\omega_{\gamma})\\
f_{\boldsymbol{p}}(1-f_{\boldsymbol{p-q}})(N_{\gamma,\boldsymbol{q}}+1)\delta(E_{\boldsymbol{p-q}}-E_{\boldsymbol{p}}+\hbar\omega_{\gamma})
\end{array}\right\} \ ,\label{eq:PhononEmissionAbsorption}
\end{equation}
where $M_{\gamma,\boldsymbol{q}}$ is the electron-phonon coupling
coefficient, $f_{\boldsymbol{p}}=\{\exp[\beta(E_{\boldsymbol{p}}-\mu)]+1\}^{-1}$
is the Fermi-Dirac occupation factor for the electronic state $\boldsymbol{p}$,
$N_{\gamma,\boldsymbol{q}}$ is the Bose-Einstein occupation factor
$N(\omega_{\gamma},T)=[\exp(\beta\hbar\omega_{\gamma})-1]^{-1}$,
and $E_{\boldsymbol{p}}=\hbar^{2}p^{2}/(2m_{e})$ and $\omega_{\gamma}$
represent the electron and phonon energy, respectively. The terms
$\beta=(k_{B}T)^{-1}$ and $\mu=\beta^{-1}\ln[\exp(\frac{2\pi\hbar^{2}n\beta}{g_{s}g_{v}m_{e}})-1]$
represent, respectively, the inverse temperature and chemical potential,
where $T$ is the temperature, $n$ is the electron density, $k_{B}$
is the Boltzmann constant, and $\hbar$ is the reduced Planck constant.

Given Eq.~(\ref{eq:PhononEmissionAbsorption}), the sums in Eq.~(\ref{eq:PhononRate})
can be written as 
\[
\sum_{\boldsymbol{p}}\left\{ \begin{array}{c}
W_{\gamma,\boldsymbol{p}\rightarrow\boldsymbol{p+q}}^{(\text{abs})}\\
W_{\gamma,\boldsymbol{p}\rightarrow\boldsymbol{p-q}}^{(\text{ems})}
\end{array}\right\} =\frac{2\pi\Omega|M_{\gamma,\boldsymbol{q}}|^{2}}{g_{s}g_{v}\hbar}\left\{ \begin{array}{c}
N_{\gamma,\boldsymbol{q}}S_{0}(\boldsymbol{q},\omega_{\gamma})\\
(N_{\gamma,\boldsymbol{q}}+1)S_{0}(-\boldsymbol{q},-\omega_{\gamma})
\end{array}\right\} \ ,
\]
where $\Omega$ is the area of the TMD-substrate interface and $S_{0}(\boldsymbol{q},\omega)$
is the dynamic structure factor of the 2DEG in the random phase approximation~\citep{GDMahan:Book00_Many},
i.e., $S_{0}(\boldsymbol{q},\omega)=\frac{g_{s}g_{v}}{\Omega}\sum_{\boldsymbol{p}}f_{\boldsymbol{p}}(1-f_{\boldsymbol{p+q}})\delta(E_{\boldsymbol{p+q}}-E_{\boldsymbol{p}}-\hbar\omega)$
which simplifies to 
\begin{align*}
S_{0}(\boldsymbol{q},\omega) & =\frac{g_{s}g_{v}[N(\omega,T)+1]}{\Omega\pi}\lim_{\eta\rightarrow0^{+}}\text{Im}\sum_{\boldsymbol{p}}\frac{f_{\boldsymbol{p+q}}-f_{\boldsymbol{p}}}{E_{\boldsymbol{p+q}}-E_{\boldsymbol{p}}-\hbar\omega+i\eta}\\
 & =\frac{N(\omega,T)+1}{\pi}\text{Im}\mathcal{P}(\boldsymbol{q},\omega;\mu,T)
\end{align*}
and 
\begin{equation}
\mathcal{P}(\boldsymbol{q},\omega;\mu,T)=\frac{g_{s}g_{v}}{\Omega}\lim_{\eta\rightarrow0^{+}}\sum_{\boldsymbol{p}}\frac{f_{\boldsymbol{p+q}}-f_{\boldsymbol{p}}}{\hbar\omega-E_{\boldsymbol{p+q}}+E_{\boldsymbol{p}}+i\eta}\ .\label{eq:2DEGPolarization}
\end{equation}
Equation~(\ref{eq:2DEGPolarization}) describes the finite-temperature
2DEG polarizability and can be written as~\citep{PFMaldague:SurfSci78_Many}
\[
\mathcal{P}(\boldsymbol{q},\omega;\mu,T_{\text{el}})=\int_{0}^{\infty}d\mu^{\prime}\frac{\mathcal{P}(\boldsymbol{q},\omega;\mu^{\prime},0)}{4k_{B}T_{\text{el}}\cosh^{2}(\frac{\mu-\mu^{\prime}}{2k_{B}T_{\text{el}}})}
\]
where $T_{\text{el}}$ is the electronic temperature and for $z_{\pm}=\hbar\omega\pm\frac{\hbar^{2}q^{2}}{2m_{e}}$
and $u=\frac{2\mu\hbar^{2}q^{2}}{m_{e}}$, its exact expression at
$T_{\text{el}}=0$ is~\citep{PFMaldague:SurfSci78_Many} 
\begin{align*}
\mathcal{P}(\boldsymbol{q},\omega;\mu,0) & =\frac{g_{s}g_{v}m_{e}}{2\pi\hbar^{2}}+\frac{g_{s}g_{v}m_{e}^{2}}{2\pi\hbar^{4}q^{2}}\left\{ \text{sgn}(z_{+})\Theta(z_{+}^{2}-u)\sqrt{z_{+}^{2}-u}-\text{sgn}(z_{-})\Theta(z_{-}^{2}-u)\sqrt{z_{-}^{2}-u}\right\} \\
 & +i\frac{g_{s}g_{v}m_{e}^{2}}{2\pi\hbar^{4}q^{2}}\left\{ \Theta(u-z_{+}^{2})\sqrt{u-z_{+}^{2}}-\Theta(u-z_{-}^{2})\sqrt{u-z_{-}^{2}}\right\} \ .
\end{align*}

\subsection{Electron-phonon interaction and screening}

The electron-phonon coupling coefficient $M_{\gamma,\boldsymbol{q}}$
in Eq.~(\ref{eq:PhononEmissionAbsorption}) is~\citep{AKonar:PRB10_Effect}
\begin{equation}
M_{\gamma,\boldsymbol{q}}=\left[\frac{e^{2}\hbar\omega_{\gamma}\exp(-2qd)}{\Omega q\varepsilon(q)}\left(\frac{1}{\epsilon_{\gamma,\text{hi}}}-\frac{1}{\epsilon_{\gamma,\text{lo}}}\right)\right]^{1/2}\ ,\label{eq:ElectronPhononCoupling}
\end{equation}
where $e$ is the electron charge. The expressions for $\epsilon_{\text{SO1},\text{hi}}$,
$\epsilon_{\text{SO1},\text{lo}}$, $\epsilon_{\text{SO2},\text{hi}}$
and $\epsilon_{\text{SO2},\text{lo}}$ in Eq.~(\ref{eq:ElectronPhononCoupling})
are 
\begin{align}
\epsilon_{\text{SO1},\text{hi}} & =\frac{1}{2}\left[\epsilon_{\text{sub}}^{\infty}\left(\frac{\omega_{\text{LO2}}^{2}-\omega_{\text{SO1}}^{2}}{\omega_{\text{TO2}}^{2}-\omega_{\text{SO1}}^{2}}\right)+\epsilon_{0}\right]\nonumber \\
\epsilon_{\text{SO1},\text{lo}} & =\frac{1}{2}\left[\epsilon_{\text{sub}}^{\infty}\left(\frac{\omega_{\text{LO1}}^{2}}{\omega_{\text{TO1}}^{2}}\right)\left(\frac{\omega_{\text{LO2}}^{2}-\omega_{\text{SO1}}^{2}}{\omega_{\text{TO2}}^{2}-\omega_{\text{SO1}}^{2}}\right)+\epsilon_{0}\right]\label{eq:DielectricFunctions}\\
\epsilon_{\text{SO2},\text{hi}} & =\frac{1}{2}\left[\epsilon_{\text{sub}}^{\infty}\left(\frac{\omega_{\text{LO1}}^{2}-\omega_{\text{SO2}}^{2}}{\omega_{\text{TO1}}^{2}-\omega_{\text{SO2}}^{2}}\right)+\epsilon_{0}\right]\nonumber \\
\epsilon_{\text{SO2},\text{lo}} & =\frac{1}{2}\left[\epsilon_{\text{sub}}^{\infty}\left(\frac{\omega_{\text{LO2}}^{2}}{\omega_{\text{TO2}}^{2}}\right)\left(\frac{\omega_{\text{LO1}}^{2}-\omega_{\text{SO2}}^{2}}{\omega_{\text{TO1}}^{2}-\omega_{\text{SO2}}^{2}}\right)+\epsilon_{0}\right]\nonumber 
\end{align}
where $\epsilon_{0}$ is the permittivity of vacuum, and the screening
function $\varepsilon(q)$ is given by~\citep{ZYOng:PRB12_Charged}
\begin{equation}
\varepsilon(q)^{-1}=1+\frac{e^{2}\text{Re}\mathcal{P}(\boldsymbol{q},0;\mu,T)}{2\epsilon_{0}q}\left[1-\frac{\epsilon_{\text{sub}}^{\infty}-\epsilon_{0}}{\epsilon_{\text{sub}}^{\infty}+\epsilon_{0}}\exp(-2qd)\right]\ .\label{eq:ScreeningFunction}
\end{equation}
As expected, Eq.~(\ref{eq:ScreeningFunction}) becomes $\varepsilon(q)^{-1}=1+e^{2}\text{Re}\mathcal{P}(\boldsymbol{q},0;\mu,T)/(2\epsilon_{0}q)$
in the $d\rightarrow\infty$ (\emph{i.e.} no substrate) limit and
$\varepsilon(q)^{-1}=1+e^{2}\text{Re}\mathcal{P}(\boldsymbol{q},0;\mu,T)/[(\epsilon_{\text{sub}}^{\infty}+\epsilon_{0})q]$
in the $d=0$ limit (\emph{i.e.} no gap between the TMD and substrate).
The screening function reduces the bare electron-phonon strength especially
in the short wavelength ($q\rightarrow\infty$) limit and originates
from the polarization of the mobile electrons in the 2DEG to screen
the surface electric field generated by the substrate. To obtain the
bare electron-phonon strength, we can set $\varepsilon(q)^{-1}=1$.

\subsection{Electronic thermal boundary conductance}

Using the identities $S_{0}(\boldsymbol{q},\omega)=e^{\beta\hbar\omega}S_{0}(\boldsymbol{q},-\omega)$
and $S_{0}(\boldsymbol{q},\omega)=S_{0}(-\boldsymbol{q},\omega)$,
Eq.~(\ref{eq:PhononRate}) simplifies to 
\[
\frac{dN_{\gamma,\boldsymbol{q}}}{dt}=-\frac{2\pi\Omega|M_{\gamma,\boldsymbol{q}}|^{2}}{\hbar}S_{0}(\boldsymbol{q},\omega_{\gamma})\frac{N_{\gamma,\boldsymbol{q}}(T_{\text{sub}})-N_{\gamma,\boldsymbol{q}}^{\text{Eq}}}{N_{\gamma,\boldsymbol{q}}^{\text{Eq}}+1}\ 
\]
where $N_{\gamma,\boldsymbol{q}}^{\text{Eq}}$ is the Bose-Einstein
occupation factor at the TMD temperature. Given the SO phonon energy
density $\varrho(T_{\text{sub}})=\frac{1}{\Omega}\sum_{\gamma,\boldsymbol{q}}\hbar\omega_{\gamma}N_{\gamma,\boldsymbol{q}}$,
we can write the electronic TBC as the derivative of the rate of change
of $\varrho(T_{\text{sub}})$ with respect to $T_{\text{sub}}$, i.e.,
\[
G_{\text{el}}=-\frac{d}{dT_{\text{sub}}}\left[\frac{d\varrho(T_{\text{sub}})}{dt}\right]_{T_{\text{sub}}=T}=\sum_{\gamma,\boldsymbol{q}}\frac{2\pi\hbar\omega_{\gamma}^{2}}{k_{B}T^{2}}|M_{\gamma,\boldsymbol{q}}|^{2}S_{0}(\boldsymbol{q},\omega_{\gamma})N_{\gamma,\boldsymbol{q}}^{\text{Eq}}\ .
\]
Hence, the final expression for $G_{\text{el}}$ is 
\begin{equation}
G_{\text{el}}=\sum_{\gamma,\boldsymbol{q}}\frac{2\hbar\omega_{\gamma}^{2}}{k_{B}T^{2}}|M_{\gamma,\boldsymbol{q}}|^{2}(N_{\gamma,\boldsymbol{q}}^{\text{Eq}}+1)N_{\gamma,\boldsymbol{q}}^{\text{Eq}}\text{Im}\mathcal{P}(\boldsymbol{q},\omega_{\gamma};\mu,T)\ ,\label{eq:ThermalBoundaryConductance}
\end{equation}
which, for the ease of numerical evaluation, we can rewrite as a multivariable
integral 
\begin{equation}
G_{\text{el}}=\sum_{\gamma=\text{SO1},\text{SO2}}\mathcal{F}_{\gamma}\int_{0}^{q_{\text{max}}}dq\ \frac{\exp(-2qd)}{\varepsilon(q)}\int_{0}^{\infty}d\mu^{\prime}\frac{\text{Im}\mathcal{P}(\boldsymbol{q},\omega_{\gamma};\mu^{\prime},0)}{4k_{B}T\cosh^{2}(\frac{\mu-\mu^{\prime}}{2k_{B}T})}\label{eq:TBCExpression}
\end{equation}
where $q_{\text{max}}$ is the maximum wave vector, which we can set
as $q_{\text{max}}=2\pi/\sqrt{A}\approx2\times10^{10}$ m$^{-1}$
($A$ is the unit cell area of the TMD), and 
\begin{equation}
\mathcal{F}_{\gamma}=\frac{e^{2}\hbar^{2}\omega_{\gamma}^{3}}{\pi k_{B}T^{2}}(N_{\gamma}^{\text{Eq}}+1)N_{\gamma}^{\text{Eq}}\left(\frac{1}{\epsilon_{\gamma,\text{hi}}}-\frac{1}{\epsilon_{\gamma,\text{lo}}}\right)\ .\label{eq:FCoefficient}
\end{equation}

We use the simulation parameters from Table~\ref{Tab:RemotePhononParameters}
in our calculations. The TMD-substrate gap size $d$ is a parameter
in a continuum model in which we treat the substrate as a continuous
dielectric solid with a well-defined surface at which the dielectric
function changes discontinuously from $\epsilon_{\text{sub}}(\omega)$
to $\epsilon_{0}$. However, there is no well-defined, exact point
in atomistic models where we can say that the dielectric function
changes suddenly. Hence, the precise value of $d$ cannot be determined
from atomistic models. Moreover, the value of $d$ estimated from
atomistic models can vary with the chemical configuration of the substrate
surface. Thus, given the uncertainty in $d$, we look at the range
of $G_{\text{el}}$ values for different values of $d$ instead of
computing $G_{\text{el}}$ for a single fixed $d$ value.

We estimate $d=0.3$ nm for the TMD-SiO$_{2}$ interfaces from the
position of the highest substrate atom to the lowest TMD atom in the
TMD-SiO$_{2}$ supercells optimized with density functional theory
(DFT) calculations, the details of which are given in the Appendix~\ref{sec:DFT_details}.
We also assume $d=0.3$ nm for the TMD-HfO$_{2}$ and TMD-Al$_{2}$O$_{3}$
interfaces for convenience. Our calculations are repeated for $d=0$
and $d=0.6$ nm, because of the strong $d$-dependence from the $e^{-2qd}$
term in Eq.~(\ref{eq:TBCExpression}), to set the upper and lower
bounds for $G_{\text{el}}$, respectively. 

\begin{table}
\begin{centering}
\begin{tabular}{|c|c|c|c|}
\hline 
Substrate & SiO$_{2}$ & Al$_{2}$O$_{3}$ & HfO$_{2}$\tabularnewline
\hline 
\hline 
$m_{e}/m_{0}$ & \multicolumn{3}{c|}{$0.31$ (WS$_{2}$) and $0.51$ (MoS$_{2}$)}\tabularnewline
\hline 
$g_{s}$ & \multicolumn{3}{c|}{$2$ (WS$_{2}$ and MoS$_{2}$)}\tabularnewline
\hline 
$g_{v}$ & \multicolumn{3}{c|}{$2$ (WS$_{2}$ and MoS$_{2}$)}\tabularnewline
\hline 
$d$ (nm) & \multicolumn{3}{c|}{$0.3$ (WS$_{2}$ and MoS$_{2}$)}\tabularnewline
\hline 
$\epsilon_{\text{sub}}^{\infty}/\epsilon_{0}$ & 2.50 & 3.20 & 5.03\tabularnewline
\hline 
$\epsilon_{\text{sub}}^{i}/\epsilon_{0}$ & 3.05 & 7.27 & 6.58\tabularnewline
\hline 
$\epsilon_{\text{sub}}^{0}/\epsilon_{0}$ & 3.90 & 12.53 & 22.00\tabularnewline
\hline 
$\omega_{\text{TO1}}$ (meV) & 55.60 & 48.18 & 12.40\tabularnewline
\hline 
$\omega_{\text{TO2}}$ (meV) & 138.10 & 71.41 & 48.35\tabularnewline
\hline 
$\omega_{\text{SO1}}$ (meV) & 60.99 & 56.08 & 21.26\tabularnewline
\hline 
$\omega_{\text{SO2}}$ (meV) & 148.97 & 110.11 & 55.08\tabularnewline
\hline 
\end{tabular}
\par\end{centering}

\caption{Remote phonon scattering simulation parameters for WS$_{2}$ and MoS$_{2}$~\citep{ZJin:PRB14_Intrinsic}.
The effective electron masses are expressed in terms of the free electron
mass $m_{0}$ and taken from Ref.~\citep{ZJin:PRB14_Intrinsic}.
The parameters $\epsilon_{\text{sub}}^{\infty}$, $\epsilon_{\text{sub}}^{i}$,
$\epsilon_{\text{sub}}^{0}$, $\omega_{\text{TO1}}$ and $\omega_{\text{TO2}}$
are taken from Ref.~\citep{ZYOng:PRB12_Theory}.}

\label{Tab:RemotePhononParameters}
\end{table}

\section{Numerical results and discussion}

To understand the experimental implications of the physics underlying
the electronic TBC, we compute the electronic TBC $G_{\text{el}}$
using Eq.~(\ref{eq:TBCExpression}) for the single-layer TMDs (MoS$_{2}$
and WS$_{2}$) and dielectric substrates (SiO$_{2}$, Al$_{2}$O$_{3}$,
and HfO$_{2}$) at different temperatures ($T$) and electron densities
($n$) since $T$ and $n$ can be varied in experiments. In field-effect
transistors, the electron density is modulated by the gate electrode
and can reach up to $n\sim10^{13}$ cm$^{-2}$~\citep{ZYu:NatCommun14_Towards,ZYu:AdvMater16_Realization}.
Intuitively, we expect $G_{\text{el}}$ to increase with $n$ as more
electron-phonon scattering events occur at higher $n$. However, at
higher $n$, the mobile electrons can also be polarized by an external
electric field and generate a polarization field that attenuates the
original external field. This effect is known as screening and it
is expected to weaken the effective electron-phonon interaction, possibly
offsetting the gain in $G_{\text{el}}$ from the greater frequency
of electron-phonon scattering. By studying the $n$ dependence of
$G_{\text{el}}$, we clarify these two competing effects on $G_{\text{el}}$.

\subsection{$G_{\text{el}}$ dependence on electron screening}

It is shown in Ref.~\citep{ZYOng:PRB13_Signatures} that $G_{\text{el}}$
in single-layer graphene can vary substantially depending on the form
of the electron-phonon interaction $M_{\gamma,\boldsymbol{q}}$ in
Eq.~(\ref{eq:ElectronPhononCoupling}), with $G_{\text{el}}$ for
the bare or unscreened electron-phonon interaction (i.e. the unscreened
$G_{\text{el}}$) up to two orders of magnitude larger than $G_{\text{el}}$
for the weaker screened electron-phonon interaction (i.e. the screened
$G_{\text{el}}$). Experimental TBC data from Koh and co-workers for
the graphene-SiO$_{2}$ interface~\citep{YKoh:NL16_Role} suggest
that changes in the TBC are only weakly dependent on the electron
density $n$, consistent with the screened $M_{\gamma,\boldsymbol{q}}$.
On the other hand, theories involving unscreened remote phonon interaction
have been used to model heat dissipation in carbon nanotubes~\citep{SRotkin:NL09_Essential}.
To settle this question for TMDs and to determine the significance
of screening for $G_{\text{el}}$ in TMDs, we compute the unscreened
and screened $G_{\text{el}}$ using Eq.~(\ref{eq:TBCExpression}).
For screened interactions, we use Eq.~(\ref{eq:ScreeningFunction})
while for unscreened interactions, we set $\varepsilon(q)^{-1}=1$
in Eq.~(\ref{eq:TBCExpression}).

\begin{figure}
\begin{centering}
\includegraphics[scale=0.5]{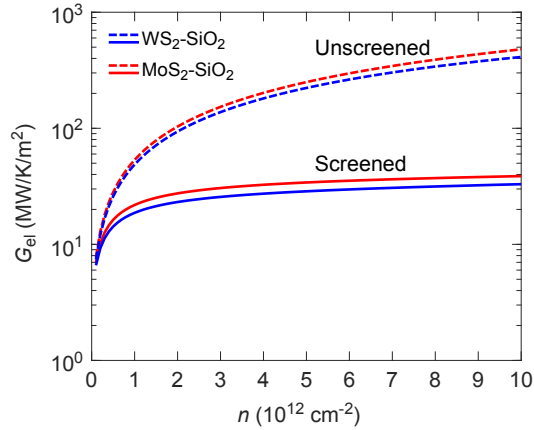}
\par\end{centering}
\caption{Electron density dependence of $G_{\text{el}}$ for the WS$_{2}$-SiO$_{2}$
(blue lines) and MoS$_{2}$-SiO$_{2}$ (red lines) interfaces with
screened (solid lines) and unscreened (dashed lines) interaction at
$T=300$ K.}
\label{fig:UnscreenedVsScreenedInteraction}
\end{figure}

Figure~\ref{fig:UnscreenedVsScreenedInteraction} shows $G_{\text{el}}$
as a function of $n$ at $T=300$ K for the MoS$_{2}$-SiO$_{2}$
and WS$_{2}$-SiO$_{2}$ interfaces. For the MoS$_{2}$-SiO$_{2}$
interface, the unscreened $G_{\text{el}}$ shows an order of magnitude
rise from $53$ to $479$ MW/K/m$^{2}$ as $n$ increases from $10^{12}$
to $10^{13}$ cm$^{-2}$ while the screened $G_{\text{el}}$ increases
from $22$ to $39$ MW/K/m$^{2}$. This difference in the dependence
on $n$ illustrates the significance of screening for the TBC. At
low $n$ where screening is weak, the unscreened and screened $G_{\text{el}}$
values are close but at high $n$, the screening of $M_{\gamma,\boldsymbol{q}}$
becomes more significant and the screened $G_{\text{el}}$ diverges
from the unscreened $G_{\text{el}}$. We surmise that the large unscreened
$G_{\text{el}}$ increase is unlikely to be correct given the much
smaller MoS$_{2}$-SiO$_{2}$ TBC observed in Refs.~\citep{EYalon:NL17_Energy,EYalon:ACSApplMaterInterf17_Temperature}.
Thus, electron screening must be taken into consideration in order
to model electronic heat transfer at TMD-substrate interfaces. Nonetheless,
even with the effects of screening, the $n$ dependence of $G_{\text{el}}$
is still significant and should be detectable in a MoS$_{2}$ or WS$_{2}$
field-effect transistor on a typical Si/SiO$_{2}$ substrate.

We observe that $G_{\text{el}}$ is larger for MoS$_{2}$ than for
WS$_{2}$. At $n=10^{12}$ cm$^{-2}$, the unscreened $G_{\text{el}}$
is $53$ and $49$ MW/K/m$^{2}$ for the MoS$_{2}$-SiO$_{2}$ and
WS$_{2}$-SiO$_{2}$ interface, respectively~\citep{ZYOng:PRB13_Signatures},
i.e., $G_{\text{el}}$ is $\sim10$ percent higher for MoS$_{2}$.
This is due to the higher electron density of states for MoS$_{2}$,
which is a constant proportional to the electron effective mass $m_{e}$~\citep{JHDavies:Book98_Physics}
and corresponds to a greater amount of scattering. The relative difference
is further enhanced for the screened $G_{\text{el}}$, with the screened
$G_{\text{el}}$ for MoS$_{2}$ about $\sim17$ percent higher, indicating
that a higher $m_{e}$ also leads to weaker screening of the electron-phonon
interaction. This is because at a higher density of states, the greater
availability of states lowers the chemical potential which in turn
decreases the screening strength. This combined effect of a lower
density of states and stronger screening also explains why $G_{\text{el}}$
is significantly lower for graphene. At $n=10^{12}$ cm$^{-2}$, the
unscreened $G_{\text{el}}$ is $53$ and $15$ MW/K/m$^{2}$ for the
MoS$_{2}$-SiO$_{2}$ and graphene-SiO$_{2}$ interface, respectively~\citep{ZYOng:PRB13_Signatures}.
Unlike the 2DEG in MoS$_{2}$ or WS$_{2}$ which has a parabolic dispersion
($E\propto k^{2}$), the 2DEG in graphene has a linear dispersion
($E\propto k$) akin to a massless Dirac particle. Thus, the electronic
density of states in graphene scales as $E$ near the Dirac point,
which results in a reduced amount of scattering and stronger screening
of the electron-phonon coupling, leading to a smaller $G_{\text{el}}$.

\subsection{$G_{\text{el}}$ dependence on substrate dielectric material, electron
density and temperature}

\begin{figure}
\centering{}\includegraphics[scale=0.5]{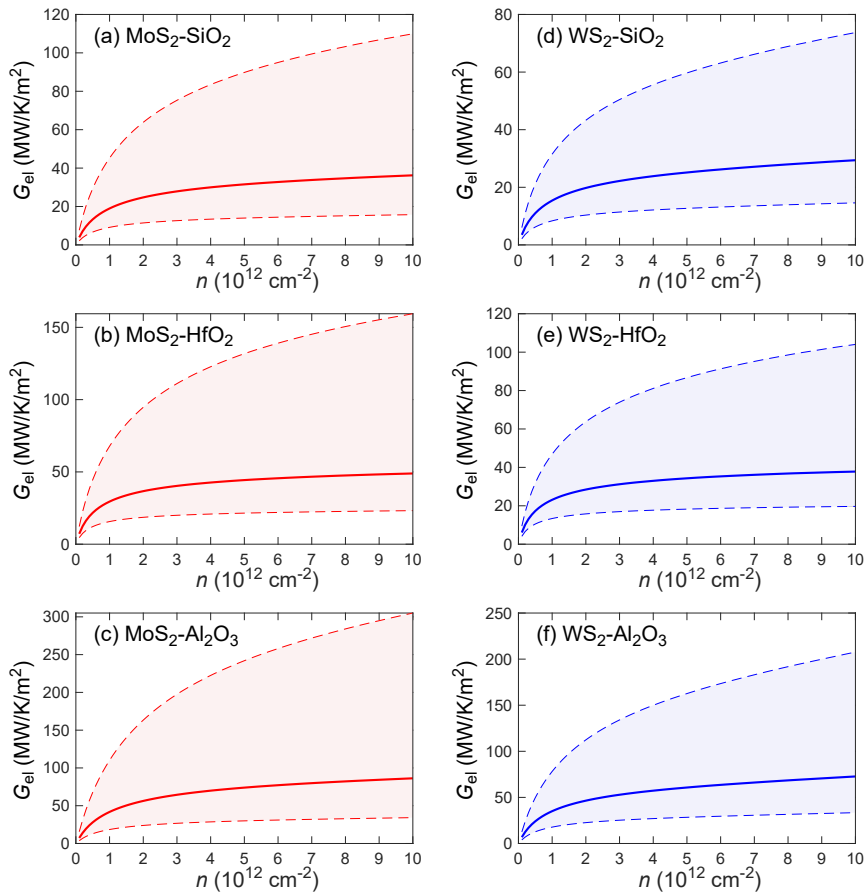}\caption{Electron density dependence of $G_{\text{el}}$ for the TMD-substrate
interface with different TMDs (WS$_{2}$ and MoS$_{2}$) and substrates
(SiO$_{2}$, HfO$_{2}$ and Al$_{2}$O$_{3}$) at $T=300$ K over
the range of $n=0.1$ to $10\times10^{12}$ cm$^{-2}$ for $d=0.3$
nm (solid line). The upper and lower bounds in each panel correspond
to $G_{\text{el}}$ for $d=0$ and $d=0.6$ nm (dashed lines).}
\label{fig:ElectronicTBCDifferentSubstrates}
\end{figure}

Figures~\ref{fig:ElectronicTBCDifferentSubstrates}(a) to \ref{fig:ElectronicTBCDifferentSubstrates}(f)
show the electronic TBC from Eq.~(\ref{eq:TBCExpression}) as a function
of the electron density $n$ at $T=300$ K for different substrates
(SiO$_{2}$, HfO$_{2}$, and Al$_{2}$O$_{3}$) and a TMD-substrate
distance of $d=0.3$ nm. The lower and upper bounds for $G_{\text{el}}$,
corresponding to $d=0.6$ and $0.0$ nm, respectively, are also shown.
As noted earlier, $G_{\text{el}}$ increases monotonically from 0
with $n$ because of the higher rate of inelastic scattering events
at higher $n$. However, the rate of increase of $G_{\text{el}}$
with respect to $n$ decreases at higher electron densities because
the higher rate of inelastic scattering events is offset partially
by the greater electron screening which weakens the electron-phonon
coupling. We find that $G_{\text{el}}$ is higher for MoS$_{2}$ than
WS$_{2}$ with every substrate and that Al$_{2}$O$_{3}$ is the substrate
with the highest $G_{\text{el}}$ because it has the highest bare
electron-phonon coupling strength which we can characterize by the
dimensionless parameter $\mathcal{C}=\epsilon_{0}/\epsilon_{\text{sub}}^{0}-\epsilon_{0}/\epsilon_{\text{sub}}^{\infty}$
($\mathcal{C}=0.147$, $0.153$ and $0.233$ for SiO$_{2}$, HfO$_{2}$
and Al$_{2}$O$_{3}$, respectively). The higher $G_{\text{el}}$
for MoS$_{2}$ can be explained by its greater electron effective
mass which is proportional to the electron density of states~\citep{JHDavies:Book98_Physics}.

By increasing the electron density to $n=10^{13}$ cm$^{-2}$ which
can be attained in MoS$_{2}$ field-effect transistors~\citep{ZYu:NatCommun14_Towards,ZYu:AdvMater16_Realization},
$G_{\text{el}}$ can reach a maximum of $85$, $48$ and $39$ MW/K/m$^{2}$
in Al$_{2}$O$_{3}$, HfO$_{2}$ and SiO$_{2}$, respectively. This
figure represents the gate-tunable component of the overall TBC, which
is the sum of $G_{\text{el}}$ and $G_{\text{ph}}$ the phononic TBC
component, and suggests that the gate-tunable $G_{\text{el}}$ is
most pronounced and easily observed for the TMD-Al$_{2}$O$_{3}$
interface at room temperature since $G_{\text{el}}\gg G_{\text{ph}}$
as theoretical results from Ref.~\citep{CFoss:2DM19_Quantifying}
estimate that $G_{\text{ph}}<4$ MW/K/m$^{2}$ for the TMD-Al$_{2}$O$_{3}$
interface. This tunability should also be observable for the commonly
used SiO$_{2}$-supported TMD field-effect transistors because classical
MD simulation results for the TBC of the MoS$_{2}$-SiO$_{2}$ interface
($15.6$ MW/K/m$^{2}$ in Ref.~\citep{PYasaei:AdvMaterInterf17_Interfacial},
$12.2$\textendash $23.5$ MW/K/m$^{2}$ in Ref.~\citep{HFarahani:CMS18_Interfacial}
and $25.6\pm3.3$ MW/K/m$^{2}$ in Ref.~\citep{SVSuryavanshi:JAP19_Thermal})
suggest that $G_{\text{ph}}$ should be in the $10$ to $25$ MW/K/m$^{2}$
range at room temperature for the TMD-SiO$_{2}$ interface, comparable
to the change in $G_{\text{el}}$ from $0$ to $39$ MW/K/m$^{2}$
when $n$ increases from $0$ to $10^{12}$ cm$^{-2}$. The tunability
of $G_{\text{el}}$ also suggests that the gate voltage can be adjusted
to modulate cross-plane heat transfer between the TMD and the substrate.
One can exploit this effect to create gate-tunable thermal insulators
by layering the TMD with other 2D materials~\citep{SVaziri:SciAdv19_Ultrahigh}
and then using a gate metal electrode to either raise or lower the
TBC to facilitate or inhibit heat transfer at the TMD-substrate interface.

\begin{figure}
\begin{centering}
\includegraphics[scale=0.5]{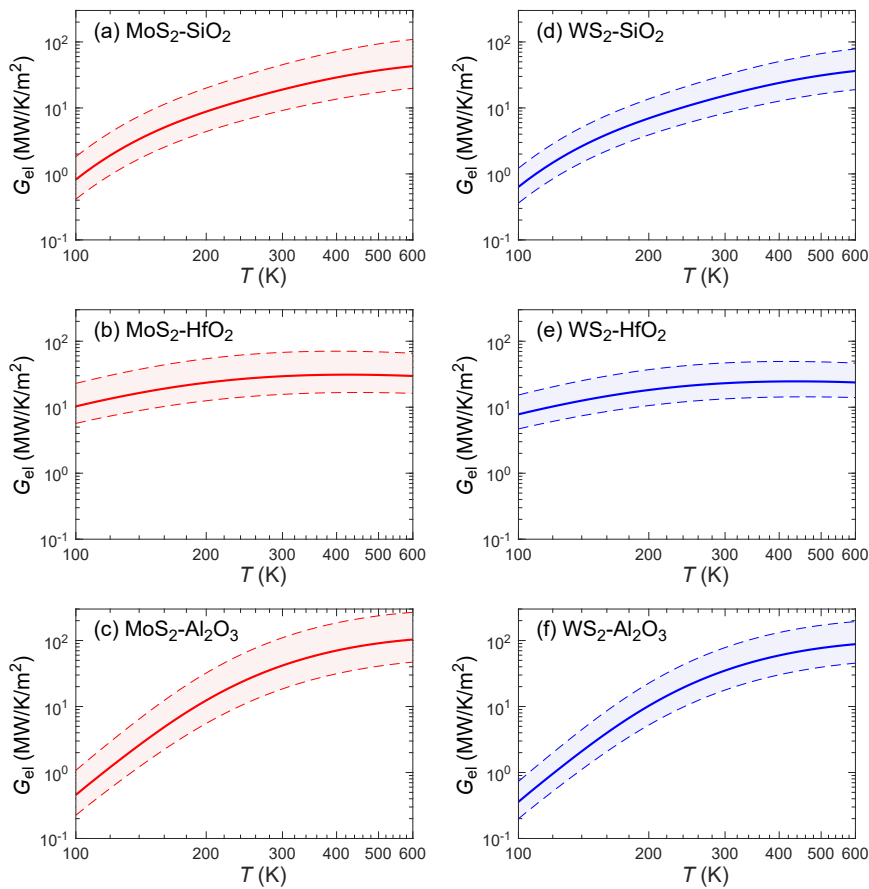}
\par\end{centering}
\caption{Temperature dependence of $G_{\text{el}}$ for the TMD-substrate interface
with different TMDs (WS$_{2}$ and MoS$_{2}$) and substrates (SiO$_{2}$,
HfO$_{2}$ and Al$_{2}$O$_{3}$) at $n=10^{12}$ cm$^{-2}$ over
the range of $T=100$ to $600$ K for $d=0.3$ nm. The upper and lower
bounds in each panel correspond to $G_{\text{el}}$ for $d=0$ and
$d=0.6$ nm.}
\label{fig:TempDependTBCDifferentSubstrates}
\end{figure}

Figures~\ref{fig:TempDependTBCDifferentSubstrates}(a) to \ref{fig:TempDependTBCDifferentSubstrates}(f)
show the temperature dependence of $G_{\text{el}}$ at $n=10^{12}$
cm$^{-2}$ for $d=0.3$ nm from $T=100$ to $600$ K, a temperature
range that is experimentally accessible. We find that $G_{\text{el}}$
increases monotonically with $T$ in this temperature range for the
TMD-SiO$_{2}$ and TMD-Al$_{2}$O$_{3}$ interfaces because the phonon
population increases with $T$. On the other hand, $G_{\text{el}}$
for the TMD-HfO$_{2}$ interface exhibits a similar trend (i.e., $dG_{\text{el}}/dT>0$)
initially but plateaus and then starts to decrease with the temperature
at higher $T$ because the expression in Eq.~(\ref{eq:FCoefficient})
scales as $\sim N_{\gamma}^{\text{Eq}}/T^{2}\propto1/T$ at higher
$T$ (i.e., $k_{B}T\gg\hbar\omega_{\text{SO1}}$) given the relatively
small SO1 phonon frequency ($\omega_{\text{SO1}}=21.26$ meV) in HfO$_{2}$.
This difference in the behavior of the temperature dependence of $G_{\text{el}}$
for different substrates can be used to distinguish the effects of
the electronic TBC. With Al$_{2}$O$_{3}$ as the substrate, we predict
that the TBC contribution from $G_{\text{el}}$ is not only large
(relative to the other substrates), but is expected to have a pronounced
temperature dependence that can be verified in experiments, while
with HfO$_{2}$, the temperature dependence is predicted to be significantly
weaker. The distinct temperature dependence of HfO$_{2}$ and Al$_{2}$O$_{3}$
suggests that different substrate materials can be used for heat-transfer
applications at different temperature regimes. HfO$_{2}$ can be used
at low temperatures ($T<200$ K) while Al$_{2}$O$_{3}$ can be used
at higher temperatures ($T>300$ K).

\section{Summary and conclusion}

In this work, we have developed a theoretical model of electronic
thermal boundary conductance (TBC) via the remote phonon scattering
of electrons in single-layer MoS$_{2}$ and WS$_{2}$ supported on
dielectric substrates. We have verified that screened electron-phonon
interactions are necessary for realistic predictions. Our model predicts
that the electronic TBC is highly dependent on the electron density
and is strongly tunable using the gate electrode in field-effect transistors,
with the strongest effect seen for Al$_{2}$O$_{3}$. We have also
identified the temperature dependence of the electronic TBC for each
dielectric substrate and the temperature regimes at which each substrate
material is more effective in interfacial heat transfer. This strong
dependence of the electronic TBC on the electron density can be exploited
for the design of novel thermal devices with gate voltage-modulated
cross-plane heat dissipative properties.

\appendix

\section{Estimation of TMD-SiO$_{2}$ gap \label{sec:DFT_details}}

The gap size ($d$) of the TMD-SiO$_{2}$ interface can be estimated
from the distance between the highest substrate atom and the lowest
TMD atom. The WS$_{2}$-SiO$_{2}$ and MoS$_{2}$-SiO$_{2}$ heterostructures
are optimized structurally within the framework of density functional
theory (DFT) using the software package VASP~\citep{Kresse:PRB96_Efficient}.
The DFT-D2 method is adopted to simulate the van der Waals interactions
across the interface while the Perdew-Burke-Ernzerhof functional is
used as the exchange correlation functional together with a cutoff
energy of 400 eV. The slab models of the heterostructures are constructed
with a vacuum layer thicker than 10 $\text{Å}$. For the TMD-SiO$_{2}$
interface, the heterostructures are constructed from a $3\times3$
supercell for the TMD (MoS$_{2}$ or WS$_{2}$) and a $2\times2$
supercell for the SiO$_{2}$ (001) surface to ensure a better lattice
match with lattice strains smaller than 2 percent. To simulate the
SiO$_{2}$ substrate, a slab model with seven Si layers is used and
the atoms in the bottom O-Si-O atomic layers are saturated with hydrogen
atoms and fixed during structural optimization. We adopt a $3\times3\times1$
Monkhorst-Pack (MP) grid for k-point sampling for the TMD-SiO$_{2}$
heterostructures. All the atomic models are fully relaxed until the
forces are smaller than 0.005 eV/$\text{Å}$.

\begin{figure}
\begin{centering}
\includegraphics[scale=0.3]{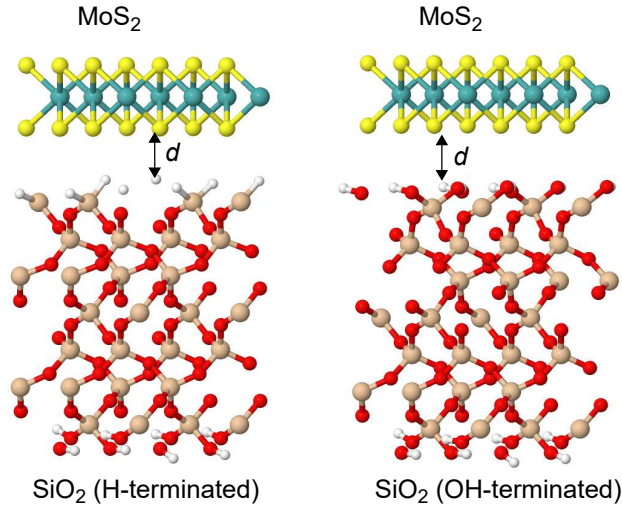}
\par\end{centering}
\caption{Side view of the MoS$_{2}$-SiO$_{2}$ heterostructures used in our
density function theory (DFT) calculations to estimate the gap size
$d$. We obtain $d=0.253$ nm (H-terminated SiO$_{2}$ surface on
the left) and $0.306$ nm (OH-terminated SiO$_{2}$ surface on the
right). The Mo, S, H, Si, and O atoms are colored in blue, yellow,
white, beige, and red, respectively. The images are generated using
Jmol~ \citep{Jmol:Jmol}.}
\label{fig:MoS2SiO2_heterostructure}
\end{figure}

Figure~\ref{fig:MoS2SiO2_heterostructure} shows the MoS$_{2}$-SiO$_{2}$
heterostructures for different SiO$_{2}$ surface configurations (H-terminated
and OH-terminated), similar to those used in Ref.~\citep{ZYOng:PRB16_Theory}.
We estimate $d=0.253$ and $0.306$ nm for the H-terminated and OH-terminated
SiO$_{2}$ surface, respectively. This shows that the estimate of
$d$ can vary with the chemical structure of the substrate surface.
Hence, it is simpler to approximate $d=0.3$ nm for the MoS$_{2}$-SiO$_{2}$
interface. Our calculations repeated for the different WS$_{2}$-SiO$_{2}$
interfacial configurations like in Fig.~\ref{fig:MoS2SiO2_heterostructure}
also suggest that $d=0.3$ nm for the WS$_{2}$-SiO$_{2}$ interface.

\begin{acknowledgments}
The authors gratefully acknowledge support from the Science and Engineering
Research Council through Grant No. 152-70-00017 and the use of computing
resources at the A{*}STAR Computational Resource Centre and National
Supercomputer Centre, Singapore. We also thank Yongqing Cai (University
of Macau) for assistance with the DFT calculations.
\end{acknowledgments}

\bibliographystyle{apsrev4-1}
\bibliography{PaperReferences}

\end{document}